\PassOptionsToPackage{table,xcdraw}{xcolor}
\documentclass[smallextended]{svjour3}       
\usepackage{tikz}

\tikzstyle{densely dashed}=[dash pattern=on 3pt off 2pt]

\smartqed  
\usepackage{graphicx}
\usepackage{subfigure}
\usepackage{amsmath}
\usepackage{balance} 
\usepackage{adjustbox}
\usepackage{multirow}
\usepackage[hidelinks]{hyperref} 
\usepackage{booktabs}
\usepackage[skins]{tcolorbox}

\usepackage{xfakebold}

\newcommand{\fbseries}{\unskip\setBold\aftergroup\unsetBold\aftergroup\ignorespaces}
\makeatletter
\newcommand{\setBoldness}[1]{\def\fake@bold{#1}}
\makeatother

\usepackage{pifont}
\newcommand{\xmark}{\ding{55}}%

\usepackage{array}
\newcommand\MyBox[2]{
  \fbox{\lower0.75cm
    \vbox to 1.7cm{\vfil
      \hbox to 1.7cm{\hfil\parbox{1.4cm}{#1\\#2}\hfil}
      \vfil}%
  }%
}

\definecolor{formalshade}{rgb}{0.93,0.93,0.93}
\newcommand{\gray}{\cellcolor{lightgray}}
\newcommand{\red}{\cellcolor[HTML]{FFCCC9}}

\begin{document}

\newenvironment{RQ}{\vspace{2mm}\begin{tcolorbox}[enhanced,width=4.7in,size=fbox,fontupper=\normalsize,colback=blue!5,drop shadow southwest,sharp corners]}{\end{tcolorbox}}

\title{SparseCoder: Advancing Source Code Analysis with Sparse Attention and Learned Token Pruning}


\author{Xueqi Yang  
\and Mariusz Jakubowsk
\and Li Kang
\and Haojie Yu
\and Tim Menzies
}


\institute{Xueqi Yang, Tim Menzies \at
              Department of Computer Science, NC State University, Raleigh, USA \\
              \email{xyang37@ncsu.edu, timm@ieee.org}           
           \and
           Mariusz Jakubowsk, Li Kang, Haojie Yu \at
              Microsoft, Redmond, USA \\
    \email{mjakubow@outlook.com, kang.kelly@microsoft.com, haojieyu@microsoft.com} 
}

\date{Received: date / Accepted: date}

\maketitle

\begin{abstract}

As software projects rapidly evolve, software artifacts become more complex and defects behind get harder to identify. The emerging Transformer-based approaches, though achieving remarkable performance, struggle with long code sequences due to their self-attention mechanism, which scales quadratically with the sequence length. This paper introduces SparseCoder, an innovative approach incorporating sparse attention and learned token pruning (LTP) method (adapted from natural language processing) to address this limitation. 
Compared to previous state-of-the-art models
(CodeBERT, RoBERTa and CodeT5), 
our experiments demonstrate that SparseCoder can handle significantly longer input sequences—at least twice as long, within the limits of our hardware resources and data statistics. Additionally, SparseCoder is four times faster than other methods measured in runtime, achieving a 50\% reduction in floating point operations per second (FLOPs) with a negligible performance drop of less than 1\% compared to Transformers using sparse attention (Sparse Atten).
Plotting FLOPs of model inference against token lengths reveals that SparseCoder scales linearly, whereas other methods, including the current state-of-the-art model CodeT5, scale quadratically.
Moreover, SparseCoder enhances interpretability by visualizing non-trivial tokens layer-wise.

\end{abstract}

\section{Introduction}
\label{intro}
\label{sec:introduction}

Source code analysis lays a pivotal foundation for many industrial Software Engineering (SE) tasks, such as vulnerability detection (static~\cite{heckman2011systematic,wang2018there,yang2021learning} and dynamic~\cite{ball1999concept}), code-comment generation~\cite{hu2018deep}, code-clone detection~\cite{jiang2007deckard,white2016deep} and automated program repair~\cite{chen2019sequencer}. Traditional source code analysis studies resort to manual feature extraction which relies heavily on expert knowledge and extensive analysis to identify the features critical to elevate the model performance ~\cite{heckman2011systematic,wang2018there,marcus2001identification}.

With the development of deep learning techniques, automatic feature extraction from raw source code has demonstrated its merits in many Software Engineering tasks~\cite{chen2019literature}, e.g., source code summarization~\cite{hu2018deep,wan2018improving,zhang2020retrieval}, automated program repair~\cite{chen2019sequencer}, clone detection~\cite{li2017cclearner,ghofrani2017conceptual,tufano2018deep,white2016deep,jiang2007deckard,vaswani2017attention}, etc. 
However, due to the increasing
flexibility and complexity of software,
 hidden anomalies in software can cause faults in the software.
Given the constraints of contemporary hardware resources, current approaches can take hours to weeks to uncover the contextual information required for such kind of vulnerability analysis. For example, when we compare SparseCoder to full attention Transformers such as CodeBERT and 
RoBERTa, it can take six to nine hours to obtain results, even for just small to medium sized problems. Such slow runtimes can complicate the development and deployment of many applications including (but not limited to) the following:
\begin{itemize}
    \item \textbf{High Computational Costs:} Running LLMs with billions of parameters is computationally intensive. Efficient models can lead to cost savings in terms of both computational power and the associated energy costs, making it economically feasible for many businesses.
    \item \textbf{Lack of deployment Flexibility in Software Systems:} Highly efficient models can be deployed across diverse platforms, from cloud servers to resource-constrained edge devices. This flexibility is crucial for software that requires local processing, such as mobile apps or certain IoT solutions.
    \item \textbf{Slow Response Times in Software Applications:} Running LLMs with billions of parameters is computationally intensive. In many software applications, such as real-time communication platforms, autonomous systems, or high-frequency trading algorithms, swift response times are essential. Efficiently designed software can deliver quicker outputs, thus enhancing the user experience and the overall effectiveness of the system.
    \item \textbf{Scalability issues of Software Services:} Many businesses need to manage a high volume of user requests simultaneously. Software designed for efficiency allows for more concurrent operations on the same hardware setup, improving the scalability of services without proportionally increasing infrastructure requirements.
\end{itemize}

In recent research, there has been considerable focus on fine-tuning large language models (LLMs) for various applications, particularly within the field of software engineering. This fine-tuning process has demonstrated substantial improvements in LLM performance across several tasks, including code summarization~\cite{ahmed2022few}, code clone detection~\cite{gao2023keeping}, code generation~\cite{wang2021codet5}, automatic source code analysis~\cite{zhang2022coditt5}. Despite the promising results, the fine-tuning process can be resource-intensive and costly~\cite{chen2023longlora}, underscoring the necessity for methods that enhance the scalability and sustainability of LLM deployment in software engineering. The significance of these points extends beyond software engineering to any application involving the analysis of recursive textual structures. This encompasses not only source code analysis but also a wide range of natural language processing tasks, such as question answering, document classification, and machine translation~\cite{beltagy2020longformer}. The methodologies presented in this paper aim to address these challenges, ensuring that the utilization of LLMs remains efficient and feasible across diverse domains.

\begin{figure*}[!t]
    \centering
    \subfigure[Token Pruning on i-th full attention layer.]{\includegraphics[width=0.33\textwidth,trim = {2.5cm 8cm 40cm 6cm}, clip]{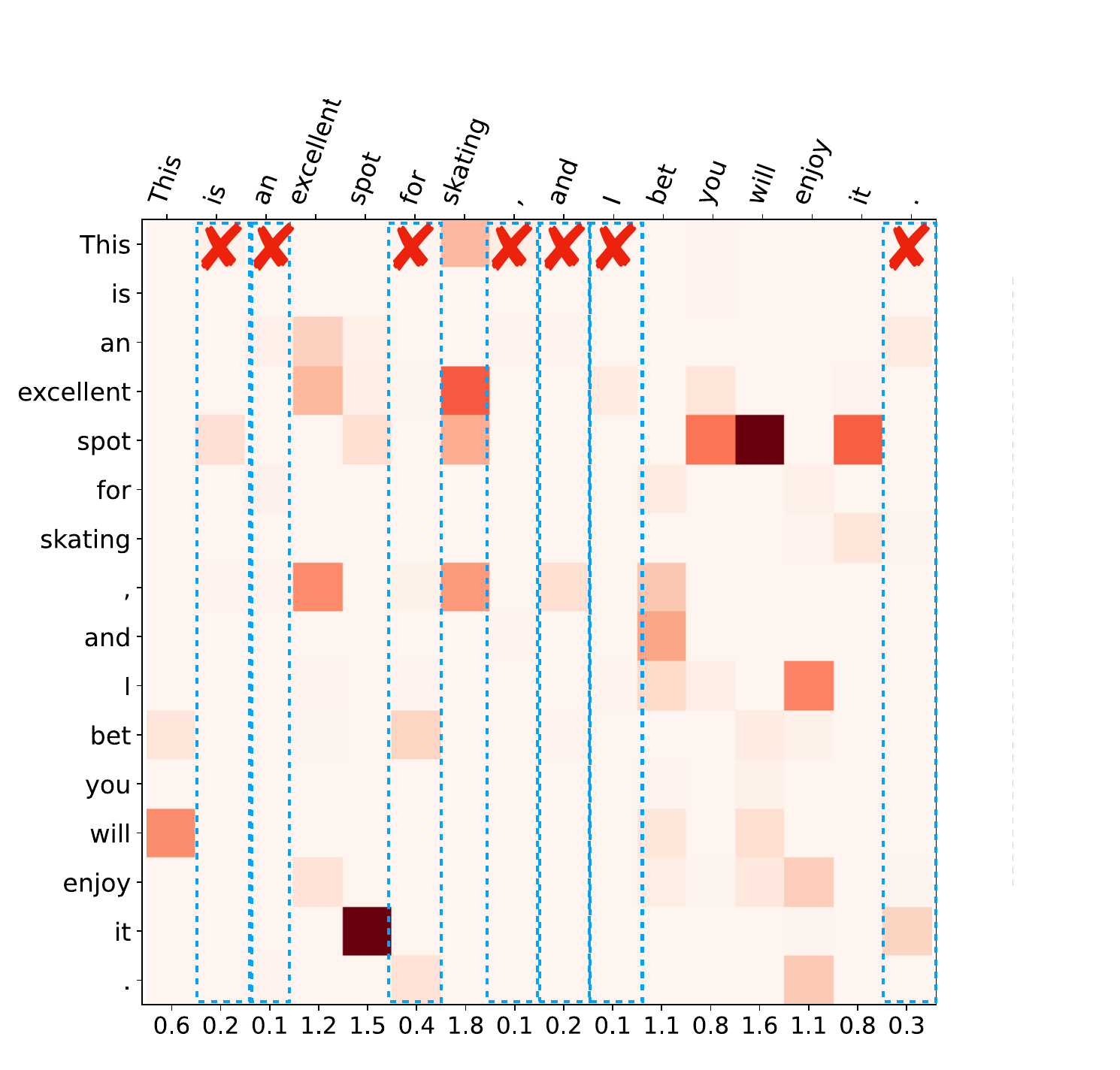}} 
    \subfigure[Token Pruning on i-th sparse attention layer.]{\includegraphics[width=0.33\textwidth,trim = {8cm 4.5cm 32cm 6cm}, clip]{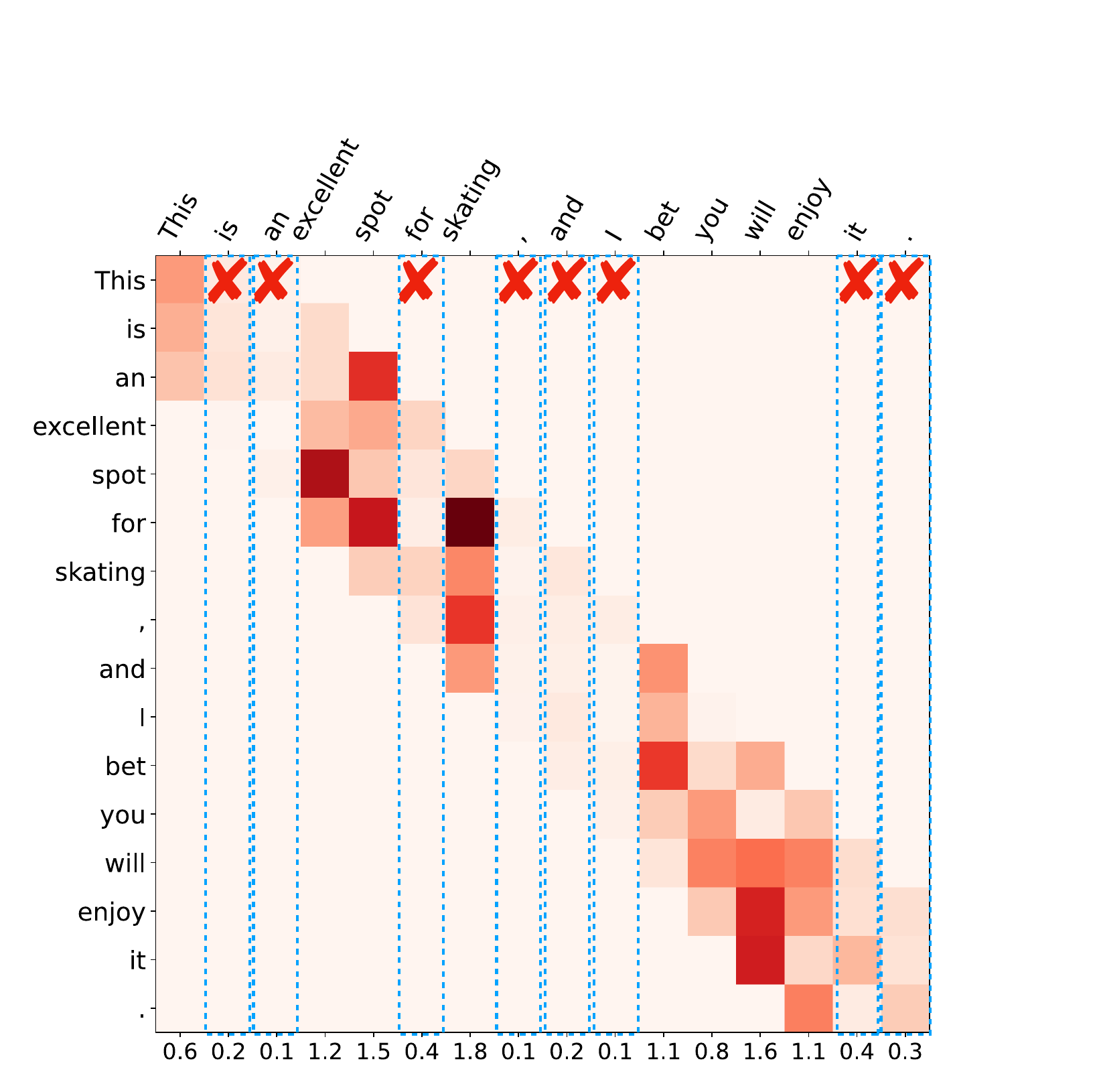}} 
    \subfigure[Visualized layer-wised token pruning.]{\includegraphics[width=0.25\textwidth,trim = {5cm 2cm 10.8cm 1cm}, clip]{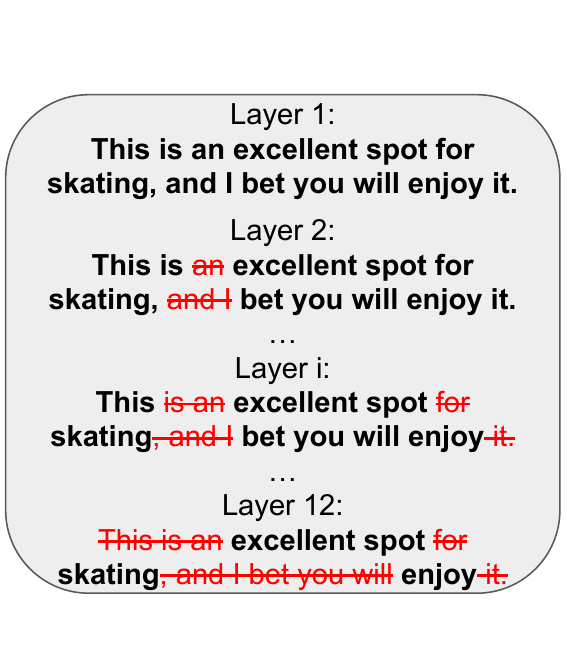}} 
    \caption{An example of SparseCoder. Token pruning on accumulative attention matrices, where full attention depicted in (a) and sparse attention visualized in (b). The accumulation is conducted vertically after row-wise softmax as formulated in Transformer models. Given a pre-defined threshold as 0.5, tokens marked with \textcolor{red}{\xmark} in both (a) and (b) denote trivial words pruned away since their accumulative attention scores fall below the threshold. (a) details the token pruning process on a single self-attention layer of Transformer model as demonstrated by Kim et al.~\cite{kim2021learned}, and (b) delves into the token pruning process within our sparse attention layer of SparseCoder, achieving greater computational efficiency through a sliding window strategy with a window size of three. Finally, (c) visualizes token pruning post-multiple attention layers, demonstrating the elimination of trivial tokens.} \label{fig:pruning}
\end{figure*}

We introduce SparseCoder, a novel integration of sparsity mechanisms and a learned token pruning algorithm, which significantly enhances the computational efficiency of defect prediction analysis. Figure~\ref{fig:pruning} demonstrates a visual representation of SparseCoder's capabilities. 
While we delve into the intricacies of this method later in the paper, it is noteworthy at this juncture to highlight that SparseCoder employs natural language processing techniques to condense extensive input sequences into more manageable lengths, specifically those with a token length greater than 512.

We raise the following research questions to assess this approach:

\begin{RQ}
{\bf RQ1.} \textbf{\textit{Scalability:}} {\em Does sparse attention mechanism advance the model scalability compared with the benchmark models (RNN-based method and Transformer with full attention, namely RoBERTa, CodeBERT and CodeT5)?}
\end{RQ}

Our experiments demonstrate that our SparseCoder extends the sequence length that models can automatically analysis. Previous Transformer models (RoBERTa, CodeBERT and CodeT5) can process max sequence length as 512, our SparseCoder achieve similar performance with much less hardware requirement and longer input sequences (over one thousands) it can analysis.
Although RNN-based models can also embed and analysis input sequences with arbitrary length, it achieves worse performance comparing with Transformer-based models.

\begin{RQ}
{\bf RQ2.} \textbf{\textit{Efficiency:}} {\em How does sparse attention mechanism advance the model efficiency compared with the benchmark models (Transformer with full attention, namely RoBERTa, CodeBERT and CodeT5)? }
\end{RQ}
 
In the experiments reported here, we observe that by incorporating sparse attention mechanism and learned token pruning algorithm, the overall model inference times were reduced about four times from over half day (for Transformer models with full attention, e.g., RoBERTa, CodeBERT and CodeT5) to a few hours (with SparseCoder) running  on the same hardware --
which in an industrial context is the difference between ``getting results tomorrow'' and ``getting results this morning''.
Further, our method scales better than prior work 
(we run in linear time while prior Transformer-based models takes quadratic time).


\begin{RQ}
{\bf RQ3.}
{\em How do the modified window size  and the token length impact the performance of the Transformer with a sparse attention mechanism?}
\end{RQ}

We demonstrate that the sparse attention mechanism can significantly improve Transformer efficiency, especially when utilizing a smaller window size. 
 Better yet, we also highlight that  overall performance increases with the growing sequence length.
\begin{RQ}
{\bf RQ4.} \textbf{\textit{Interpretability:}} {\em Can SparseCoder advance interpretability of the Transformer-based models via a token pruning algorithm?}
\end{RQ}

Our results illustrate that we can further improve the above results with SparseCoder via integrating token-pruning.\\

Our contributions can be concluded as follows:

\begin{itemize}
\item {\bf Scalability:} This paper explores an empirical study of applying a new state-of-the-art sparse attention mechanism of Transformer models in source code analysis to address the model scalability on long sequence analysis. Although in recent few years, increasing research efforts in the area of Software Engineering are put into source code analysis, our literature review demonstrates that most published works are focused on short sequence analysis on source code. And when it comes to longer and more complicated source code or SE artifacts, truncation on input sequences is required based on expert knowledge.
 \item \textbf{Efficiency}: By integrating a learned token pruning algorithm into Transformer with sparse attention, SparseCoder can adaptively prune unessential tokens layer-wised and significantly reduce the model's inference overhead from quadratic time to linear time measured in FLOPs. Our experiments shows the overall inference time of SparseCoder achieve similar performance with about four times faster inference time.
 \item \textbf{Interpretability:}
 We improve the interpretability of Transformer models with sparse attention by visualizing the important tokens after the adaptive token pruning process.
 \item {\bf An ablation study:}
 We conduct a comprehensive ablation study with different configurations of sparse attention Transformer models, which explores the impact of local information (via modified window sizes and maximum sequence lengths).

   
\end{itemize}

The rest of this paper is structured as follows. Motivation of this work is briefed in Section~\ref{sec:motivation}. Related work are introduced in Section~\ref{sec:background}. In Section~\ref{sec:method}, we illustrate the detail of our methodology. The experimental design and data curation is introduced in Section~\ref{sec:experiment}, and proposed research questions are answered in Section~\ref{sec:result}. We discussed the threats to validity and future work in Section~\ref{sec:discussion}. And finally, the conclusion is drawn in Section~\ref{sec:conclusion}.



\section{Motivation}
\label{sec:motivation}



Generalizability is a cornerstone criterion for feature engineering in software engineering domain. While extensive efforts have been put into feature selection, it remains challenging to recognize a universal feature set that generalizes well to different SE tasks. 
Moreover, it requires extra effort to integrate feature extraction into SE models, adding additional implementation complexity to the model deployment. For example, Wang et al.~\cite{wang2018there} implemented a feature extraction tool named slicer to extract all hand-crafted features (116 in total) from raw source code for vulnerability detection in Java projects. However, due to the confliction (different sets of features explored in isolated software engineering studies) and ambiguity (features with the same name indicating distinct meaning in different studies), exhaustive feature construction process (systematical literature review to collect reliable features) is required to eliminate the gap between different research studies. Moverover, as software engineering projects developed, different features could be extracted from the same project since the project pattern evolves~\cite{yang2021learning}.

Neural network-based methods~\cite{russell2018automated,kim2021learned,vaswani2017attention}, given their intrinsic advantages such as automatic feature learning, hierarchical feature extraction, and knowledge transfer, can be exceptionally beneficial. For instance, Transformer, a representative neural network architecture, can take raw code snippets as the input, process them layer by layer, and ultimately produce code analysis outcomes.
In recent years, there has been a growing interest in combining neural networks with software code analysis. However, these methods are primarily tailored for short code sequences. Previous works~\cite{hu2018deep,wan2018improving,zhang2020retrieval,wu2020code} focus on analyzing short sequences of source code, with sequence length less than 400 (most of the studied research works in software engineering domain focus on source code length lying within 200). Coinciding with this, the rise of open-source projects has made vast amounts of long code sequences readily available(\textbf{scalability issue}).

This presents an immediate demand for effective analysis of these longer code segments, a task that is notably challenging. First, as code sequences grow, software artifacts manifest increasingly complex patterns, thereby complicating code analysis. Second, processing long code sequences places a significant computational burden on Transformer architectures (\textbf{efficiency issue}). This is mainly due to their self-attention mechanism, which exhibits quadratic computational complexity relative to sequence length. Vaswani et al.~\cite{vaswani2017attention} demonstrate that the maximum sequence length that these models can analysis is 512 with given the hardware. While the extensive capability and performance of Large Language Models (LLMs) are impressive, real-world deployment often requires a balance between the number of model parameters and performance. Techniques that enhance model efficiency ensure that the benefits of LLMs can be reaped across a broader range of applications and deployment scenarios~\cite{fan2023large,hou2023large,ozkaya2023application}, e.g., deployment flexibility in software systems, enhanced response times in software applications, and efficient training in software development, etc.

A prevalent remedy involves discarding irrelevant code tokens following hardcoded rules, which means human efforts are needed to decide which segment are trivial to be discard so that the input sequence can be pruned within 512 sequence length that the model can process and analysis. Nonetheless, this method lacks flexibility and demands substantial manual effort. 
An increasing number of research studies and industrial efforts are focused on leveraging deep learning models, such as neural networks and Transformer-based models. However, these approaches often function as ``black-box'', making decisions without providing insight into the reasoning behind them. Additionally, they lack the ability to offer interpretable and transparent analysis tools for the software engineering domain, failing to highlight the code segments that contribute to their final decisions
(\textbf{interpretability issue}).

\begin{figure*}
\centerline{\includegraphics[width=4.7in,trim = {2cm 6.7cm 2cm 5.5cm}, clip]{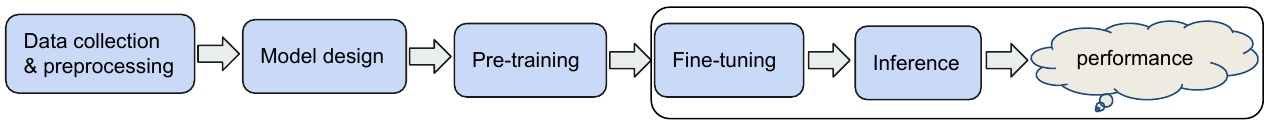}}
\caption{Pipeline of utilizing natural language models in downstream tasks. In our empirical study, only the fine-tuning and inference stages are leveraged. Fine-tuning refers to adjusting the parameters of pre-trained NLP models with a training set of the specific downstream task, and inference means evaluating the fine-tuned models on the test (new/ unseen) datasets of our downstream task.}
\label{fig:pipeline}
\end{figure*}

\section{Related Work}
\label{sec:background}

In recent years, there has been a growing interest towards integrating neural networks with software code analysis. As elaborated further,  many of these methods are primarily tailored for short code sequences -- which is
inappropriate for many industrial contexts. In many industrial contexts, certain code analysis tasks can become notably intricate due to the escalating scale of projects. To illustrate, the first author's summer internship at Google and Microsoft Research revealed the following observations:
\begin{itemize}
\item 
At Google, a software engineer may submit a single CL (change list to Google Codebase, Google3) that encompasses multiple scripts or projects. This action can produce a snapshot of a code change list spanning hundreds or thousands of lines of code.
\item 
In Microsoft's Windows Defender system, statistical results from an extensive offline PowerShell dataset in 2022 indicated that over 40\% warning messages have lengths exceeding thousands of tokens. Such lengths surpass the capabilities of conventional models.
\end{itemize}
Given the outlined challenges, the following sections of this paper delve into methods for pruning large token spaces specifically within the context of vulnerability detection in source code analysis.

\subsection{Source Code Analysis}
\label{sec:source_code_analysis}

Source code analysis serves as a critical foundation for numerous Software Engineering (SE) tasks, such as vulnerability detection (both static~\cite{heckman2011systematic,wang2018there,yang2021learning} and dynamic~\cite{ball1999concept}), source code summarization~\cite{hu2018deep}, code clone detection~\cite{jiang2007deckard,white2016deep}, and automated program repair~\cite{chen2019sequencer}, among others.

\subsubsection{Vulnerability Detection}

The Common Vulnerabilities and Exposures (CVE) organization~\footnote{\url{https://cve.mitre.org/}} published a statistical analysis that reports the tremendous growth of vulnerability numbers discovered in the last two decades, from less than 4,600 in the year 2000 to almost 20,000 currently.
These security anomalies can affect systems negatively both from financial and societal aspects, which are usually caused by subtle errors made by the programmer and even propagated promptly due to source code reuse and clones. However, most of the existing research works~\cite{ball1999concept,heckman2011systematic,wang2018there,yang2021learning} focus on short sequence analysis, mainly due to the model capacity and computing complexity. Other studies utilizing RNNs and LSTMs can address tasks with long input sequences that suffer from lack of capability to capture long-range context dependencies. Also, RNN and LSTM models depend on sequential processing, which handles an input sequence word by word. When these models encode the next word, they require the hidden state of the current word, which hinders parallel computing.

\subsubsection{Source Code Summarization}
Source code summarization, extensively studied in recent years, is to generate a short and concise natural language descriptions of source code to facilitate developer understanding and maintenance of source code.
Hu et al.~\cite{hu2018deep} propose \textit{DeepCom} via combining the natural language processing techniques, Abstract Syntax Tree, and Encoder-Decoder framework to automatically generate comments for Java methods to help developers comprehend Java programs when maintaining such projects. 
Wan et al.~\cite{wan2018improving} introduce an improved deep reinforcement learning framework by incorporating an abstract Syntax Tree structure as well as sequential content of code snippets to tackle automatic source code summarization tasks. 
Wu et al.~\cite{wu2020code} present a structure-induced Transformer model via encoding sequential code inputs with multi-view structural clues to capture various semantics of programs in source code summarization tasks.
Zhang et al.~\cite{zhang2020retrieval} take advantage of both neural and retrieval-based techniques in source code summarization by combining both the input code snippet for testing and its two most similar snippets retrieved in the training set from syntax and semantics aspects. The default input sequence length of these works is 400 or less.



\subsubsection{Code Clone Detection}
Code clone detection~\cite{li2017cclearner,ghofrani2017conceptual,tufano2018deep} is an essential task for the maintenance and evolution of software projects by evaluating the similarity of internal source code representation, such as identifier names, syntactic fragments at statement and function level, etc~\cite{white2016deep}. Recent studies on code clone detection define the four major types of clone detection, namely exact clones, renamed clones, near-miss clones and semantic clones. There also exist some other types utilized when referring to the clone relation to their experiments; e.g., structure clones and function clones. 
White et al.~\cite{white2016deep} propose learning-based code clone detection techniques by utilizing Recurrent Neural Network to automatically associate patterns extracted at the lexical level with patterns found at the syntactic level by inducing representations at different levels of granularity. Compared with a prominent structure-oriented technique, Deckard\cite{jiang2007deckard} which leverages a parsing tree instead of AST, White et al.~\cite{white2016deep} reported code clone pairs which are undetected or suboptimally reported in Deckard. Although White et al. claim that their approach can encode arbitrarily long sequences of embeddings to characterize fragments, this RNN-based method also suffers from long-range context dependencies, as most sequence transduction approaches do. It is more challenging for these models to learn long-range dependencies for longer paths between the combination of two positions in the input and output sequences~\cite{vaswani2017attention}. However, current research works mostly focus on analyzing short sequences of source code. 

\subsection{Transformers}

Transformer-based models~\cite{vaswani2017attention,jiao2019tinybert,sanh2019distilbert,brown2020language} have
achieved state-of-the-art results in sequence analysis tasks~\cite{dosovitskiy2020image,dong2018speech,chirkova2021empirical}, e.g., RoBERTa~\cite{liu2019roberta} in natural language processing and ViT~\cite{dosovitskiy2020image} in computer vision. However, it becomes more and more challenging to apply these architectures to downstream tasks efficiently, given the large model sizes, increasingly complex datasets, demand for real-time inference and limited computing resources. Most language models, not only Transformer-based models, utilize the pipeline as illustrated in Figure~\ref{fig:pipeline} to conduct classification or regression on downstream tasks. 
As illustrate in Figure~\ref{fig:pipeline}, model inference in Natural Language Processing (NLP) refers to the process of using a pre-trained and fine-tuned model to make predictions or generate outputs based on new and unseen data. This stage follows the pre-training and fine-tuning phases and involves applying the learned patterns and parameters to accomplish specific tasks, such as text classification, translation, or question answering.
Since pre-training is a time-consuming and GPU-demanding stage, our empirical study leverages only the fine-tuning and inference stages. 
Various methodologies are proposed to enhance the model efficiency during the inference stage. Pruning proposed by Lecun et al.~\cite{lecun1989optimal} is one of the popular approaches in the compression of Transformer models. Generally, by getting rid of trial or unimportant weights in neural networks, pruning can reduce inference time and memory requirement with limited performance loss by avoiding unnecessary computation with limited performance loss~\cite{treviso2022efficient}.

Previous studies~\cite{beltagy2020longformer,chen2023longlora,zaheer2020big} have shown the computation cost (e.g., memory requirement) grows quadratically with the input sequence length in the attention layers of Transformer models. This research topic has increasingly garnered attention from both the industrial sector and the research community. Several research works focus on incorporating different sparse attention mechanism into Transformer models, e.g., \textit{Longformer} (ETC)~\cite{beltagy2020longformer}, \textit{Extended Transformer Construction} (ETC)~\cite{ainslie2020etc}, BigBird~\cite{zaheer2020big},  \textit{Global Memory Augmentation for Transformers} (GMAT)~\cite{gupta2020gmat} and \textit{LongLoRA}~\cite{chen2023longlora}. 
Most of the released research papers and pre-prints are pre-trained based on long documents in NLP tasks, utilizing natural language datasets. 
Research work, \textit{CodeReviewer}~\cite{li2022codereviewer} proposed by Li et al., is pre-trained based on a large open-source dataset from Github in a code review scenario, consisting of code diff hunks and code review comments. 
In the realm of LLMs, it is imperative to augment their capability to process extended context, particularly when dealing with real-world datasets. Such an approach not only widens the model's contextual understanding but also ensures its applicability and effectiveness in addressing the nuanced requirements of practical software development scenarios.

\begin{figure}[!t]
    \centering
    \subfigure[full attention]{\includegraphics[width=0.24\textwidth]{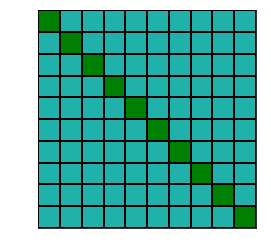}}    
    \subfigure[Local+global]
    {\includegraphics[width=0.24\textwidth]{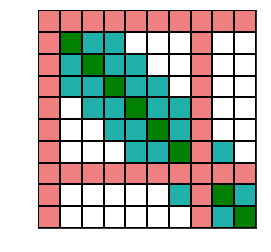}} 
    \hspace{2em}
    \subfigure[Global]{\includegraphics[width=0.086\textwidth]{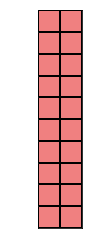}}
    \hspace{2em}
    \subfigure[Aligned local]{\includegraphics[width=0.144\textwidth]{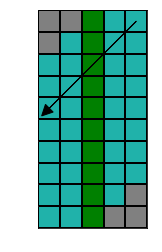}}

    \caption{Illustration of combining local and global attention mechanisms and how to efficiently store the matrix on hardware. full attention (a), local + global attention where the global attention score is marked as pink (b),  decomposing the global attention (c) and efficiently stored local attention (d).} 
\label{fig:attentions}
\end{figure}

\subsubsection{Attention Mechanisms of Transformers}

\textbf{Self-Attention:} Self-attention or full attention mechanism, is widely used in Transformer-based models. Given a sequence chunk with length $n$ in natural language, we preprocess the chunk with basic natural language preprocessors.
After conducting tokenization, each token is fully connected in the multi-head self-attention layer to make everything routed to everything. The highlighted principal diagonal means the attention of every token to itself as shown in (a) in Figure~\ref{fig:attentions}. The units are the tokens from the input sequence. 
The complexity is dominated by $O(n^2)$. To advance the model efficiency, our SparseCoder modifies the Transformer architecture by adopting the the following sparse attention mechanism.

\textbf{Local Attention:} Although the full attention mechanism is powerful as a vital attribute of the transformer-based models, these prior proposed models have a core disadvantage due to the quadratic dependency in terms of memory. 
Consequently, given the available computing resources, this approach could not process entire long input sequences (with over 512 tokens) at the same time. 
To address this limitation, there are several research works that explored the feasibility of sparse attention mechanisms to analyze longer input sequences by reducing the overall algorithm complexity.
\\
Beltagy et al.~\cite{beltagy2020longformer} propose Longformer, a modified Transformer architecture by adopting local and global attention operations that scale linearly with the sequence length, making it practical for processing long sequences. Within the framework, the local attention mechanism utilizing a sliding window scheme is the crucial component in reducing the complexity and scaling the input to long sequences.

\textbf{Global attention} remedies the attention dependency loss in the long sequence for the local attention by attending the special tokens to every other token in the input sequence. The technical details of these two attention mechanisms will be illustrated in the following subsections. 
There are several research papers also making attempts at sparse attention similar to Longformer by combining different global (e.g., random attention generated with graph) and local attention mechanisms to Transformer-based models; e.g., \textit{Extended Transformer Construction} (ETC)~\cite{ainslie2020etc}, BigBird~\cite{zaheer2020big} and \textit{Global Memory Augmentation for Transformers} (GMAT)~\cite{gupta2020gmat}. Most of them are pre-trained on long documents in NLP tasks.

\begin{figure}
 \begin{center}
\includegraphics[width=2.7in]{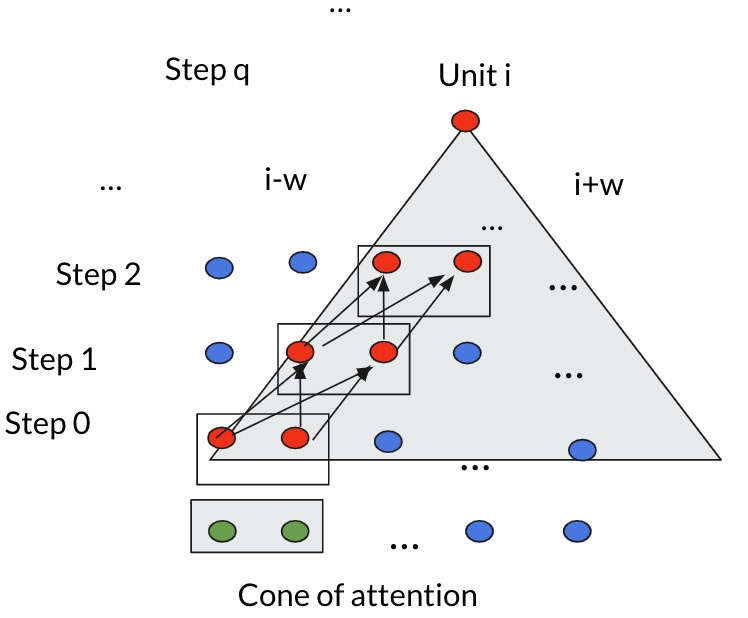} \end{center}
\caption{A demonstration of sliding window mechanism for local attention in Transformer, where the token length is \textit{n}, the window size is \textit{w} and \textit{i} is the \textit{i}-th token in the sequence.}
\label{fig:sliding_window}
\end{figure}

\subsection{Pruning}

To enhance the model efficiency,  there are generally two groups of pruning approaches based on the pruning patterns. Unstructured pruning removes less salient connections in arbitrary patterns of sparse parameters and feature maps in deep learning models. However, research works show that sparse networks with unstructured pruning do not yield significant efficiency gains when deployed on GPUs. Structured pruning removes a large part of network in structural ways, such as a layer or a channel in a CNN, or a head in a multi-head self-attention layer in the Transformer~\cite{voita2019analyzing}. However, the latter approach mainly focuses on facilitating hardware implementation instead of diving into a profound analysis of the inner characteristics of model sparsity.

A recent observation in sequence analysis is that not all tokens in the input sequences are necessary to enhance model performance, and the overall computation during model inference can be significantly reduced by removing less pertinent tokens~\cite{kim2021learned}. 
Compared with model parameter pruning, token pruning can be utilized to handle token sparsity by getting rid of less salient tokens in the importance matrix while preserving performance ~\cite{goyal2020power,kim2020length,wang2021spatten,ye2021tr,kim2021learned}. Prior studies about token pruning can be generally categorized into three families: 1) based single configuration of pruning rate, 2) top-k of the sequence length, and 3) threshold adapted on both the sequence length and the input context. Figure~\ref{fig:pruning} is the visualization of token pruning by weights in one of the full attention layers. Given an example sequence, the full attention layer attends each pair of two tokens and generates the attention score matrix. After conducting normalization on the attention matrix via a softmax function horizontally in Transformer model, the matrix can be visualized as with the heatmap. We can accumulate the attention score vertically to summarize the importance score of each token as illustrated at the bottom of Figure~\ref{fig:pruning}. Then, different pruning strategies can be applied to prune the tokens.

For the first family of token pruning, such a single configuration of pruning rate may incur over-pruning on short sequences and under-pruning on long sequences, which can damage the model performance. 
In the second family of token pruning, such as Spatten~\cite{wang2021spatten},
a proportional configuration for each layer is utilized to remove the trivial tokens and the pruning ratio is adaptively adjusted with the sequence length. However, contextual information is not considered when adjusting the pruning ratio in each layer. In the third family, Kim~\cite{kim2021learned} proposed Learned Token Pruning (LTP) by utilizing threshold-based pruning to adaptively conduct token pruning. In the fine-tuning stage, both the model parameters and the threshold are optimized based on sequence length and context. In the pruning stage, the accumulative attention score of each token is compared with a learned threshold in every layer to adaptively remove the trial tokens.\\

\section{Methodology}
\label{sec:method}

\subsection{Baselines}
\label{sec:baseline}

In this study, we leverage four prominent methodologies in NLP and source code analysis domains as baseline approaches to compare with and evaluate the efficiency of SparseCoder. And here is the recap of our baseline methods.


\subsubsection{Recurrent Neural Networks} RNNs such as LSTM~\cite{hochreiter1997long} and gated RNN~\cite{chung2014empirical}, were firmly established as prominent methods for sequence analysis, machine translation, and language modeling. These models make recurrent connections from neighboring positions of two words and generate a sequence of hidden states $s_t$ from previous hidden state $s_{t-1}$  and current position $t$.
Previous research works~\cite{white2016deep} in SE domain illustrate that RNN-based models could encode arbitrarily long sequences of embeddings to characterize code snippets. White et al.~\cite{white2016deep} show that RNN-based model can achieve State-of-the-Art performance in code clone detection. 
For the \textit{recurrent neural network (RNN)}, we leverage the bench-mark architecture \textit{Gated Recurrent Unit (GRU)}~\cite{cho2014properties} as it has advantages over \textit{long short-term memory (LSTM)} for model efficiency with less memory and faster than LSTM.
We compare the experimental results of Transformer with sparse attention to the RNN model, which is widely adopted in raw source code analysis~\cite{chen2019sequencer}.
However, the inherent sequential nature hinders the parallelization of the training process. These models tend to miss global information when it comes to long-dependency sequence analysis.

\subsubsection{RoBERTa} RoBERTa~\cite{liu2019roberta} is a replication study by Facebook AI in 2019 based on the checkpoints generated on the BERT model. This study demonstrates that the prior benchmark model BERT~\cite{devlin2018bert} is significantly undertrained. The authors showed this by implementing several simple modifications, namely training over more data with longer sequences and greater batches, removing the next sentence prediction objective and changing the masking pattern dynamically on training data. The resulting model generates competitive results on all nine natural language tasks (GLUE) compared with prior benchmark models. Several research works found RoBERTa obtaining state-of-the-art (SOTA) performance in software engineering tasks, such as code completion~\cite{ciniselli2021empirical}, vulnerability detection~\cite{do2024optimizing} and etc.

\subsubsection{CodeBERT}
CodeBERT~\cite{feng2020codebert} is the first \textit{NL-PL} Transformer-based framework pre-trained on both natural language and six programming language datasets. 
The model parameters are optimized in the pre-training process with two objectives, masked language modeling (MLM) and replaced token detection (RTD).
Although the architecture of CodeBERT is identical to RoBERTa, CodeBERT achieves benchmark results on \textit{NL-PL} tasks, such as natural language code search and code documentation generation. It's one of the most widely used Large Language Models in software engineering domain~\cite{wang2024software}.


\subsubsection{CodeT5}

CodeT5~\cite{wang2021codet5} is a large-scale language model developed by Salesforce AI Research, based on the T5 (Text-to-Text Transfer Transformer) architecture. 
It is a versatile code-aware language model pre-trained on extensive source code corpora in eight widely-used programming languages, including Python, C and Java, sourced from GitHub. 
CodeT5 is designed to assist with code-related tasks such as automatic code generation, code summarization, and code translation between different programming languages. It can understand natural language descriptions of programming tasks and generate corresponding code snippets, or vice versa. 
Recent research papers~\cite{gao2023keeping,wang2024software,wang2023rap} published in SE venues report that CodeT5 achieves SOTA performance in multiple tasks in software engineering domain, e.g., code understanding and code generation tasks, automatic program repair and etc.




\subsection{SparseCoder}

In this work, we introduce a novel framework, SparseCoder, by implementing an attention score extractor, as shown in (c) and (d) of Figure~\ref{fig:attentions}. This approach is necessary because the accumulated attention scores for each token, required by the token pruning algorithm in sparse attention, differ from those in a self-attention mechanism. The overall architecture of SparseCoder is depicted in Figure~\ref{fig:sparsecoder}.
SparseCoder can adaptively prune away unimportant tokens layer-wisely in the fine-tuning stage of Transformer models and advances the model efficiency by reducing computing overhead. More technical details about token pruning will be provided in Section~\ref{sec:pruning}. 

After conducting the layer-wised learned token pruning, both model parameters and thresholds are trained to optimize the model performance. And binarized masks are set for tokens in the hard pruning stage, where tokens with a mask as 0 are removed and a mask as 1 are kept. We can retrieve the mask information in the last (12-th) layer of SparseCoder from the neural level and visualize the input sequence after conducting token pruning. After that, the words or code snippets are restored from tokens by the mapping rules generated during the tokenization and embedding procedure.

In the following subsections, we will illustrate the two key components integrated into our SparseCoder framework:

\begin{figure*}
\centerline{\includegraphics[width=4.8in,trim = {2cm 6.7cm 3cm 3.5cm}, clip]{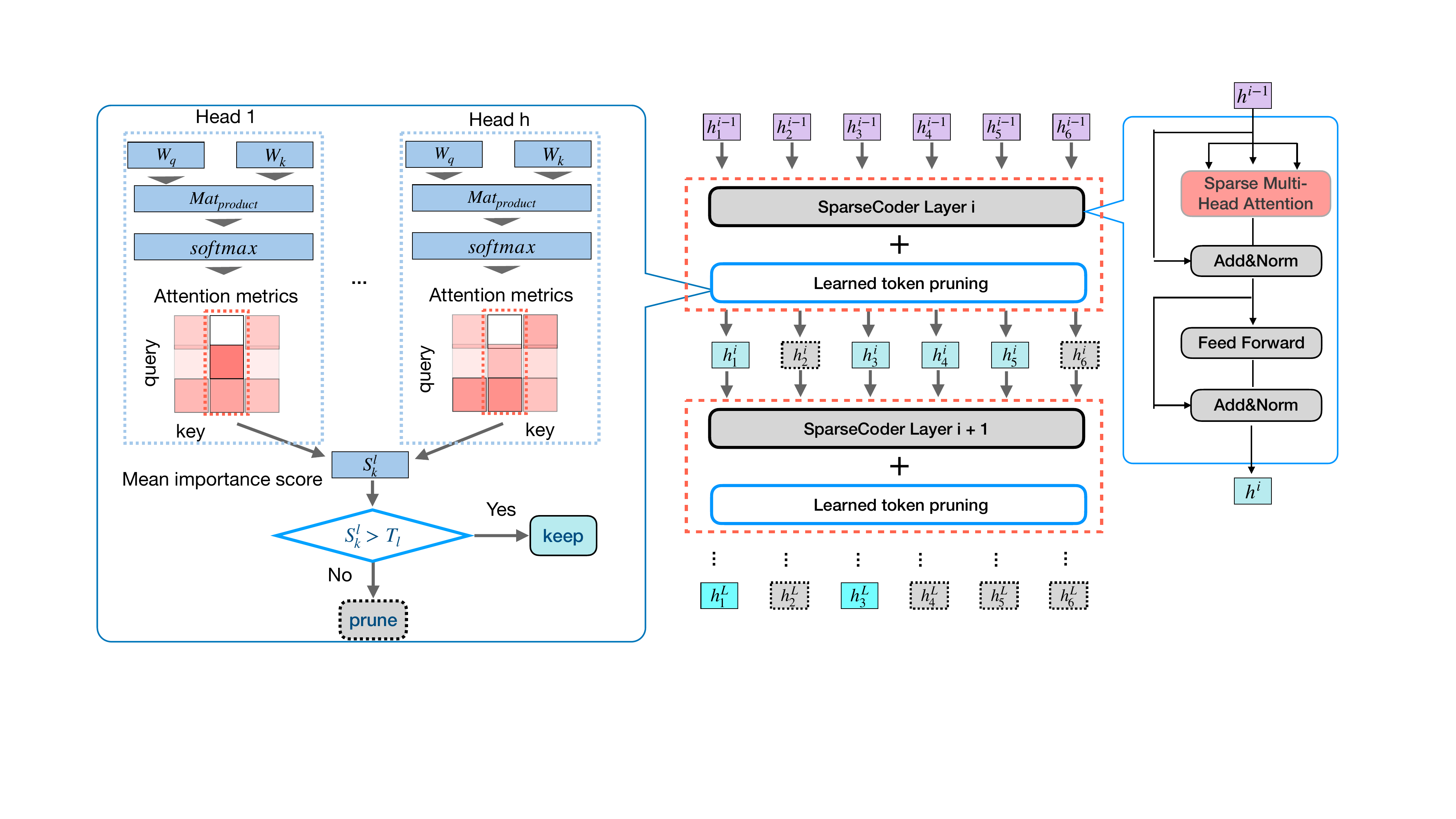}}
\caption{Demonstration of the overall structure of SparseCoder. As shown in this figure, compared with Transformer architecture, our proposed framework SparseCoder (core architecture highlighted with a \textcolor{red}{red dashed box - - - -}, which is consisted of sparse attention shown in right module and learned token pruning explained in the left module). In Transformer-based baselines (RoBERTa, CodeBERT and CodeT5), only self-attention mechanism is utilized, as shown in sub-figure (a) in Figure~\ref{fig:attentions}. While in our SparseCoder, sparse attention mechanism (For details of sparse attention, please see sub-figure (b)-(c) in Figure~\ref{fig:attentions}) can reduce the computational overhead and extend the token length that model can analysis. SparseCoder further incorporates learned token pruning (as shown in the left module) to prune away trial tokens and reduce the model inference cost.}
\label{fig:sparsecoder}
\end{figure*}

\subsubsection{Sparse Attention}





The sparse attention mechanism in SparseCoder leverages sliding window to update the attention scores of tokens in each layer, which significantly reduces the memory requirement of models. As illustrated in Figure~\ref{fig:sliding_window}, the input sequence in the example is given a sliding window with a window size of 2 for easy illustration here. 
Given a sequence of length $n$, we start from the green units from left to right; for each step from bottom to top in this example, the window is slid to get each unit covered by the window size (which is marked as red) in the current step connected with all the red units in its next step. As such, every unit is attended to by its immediate neighbors. As the window is slid across the input sequence, \textit{Unit i} in step $q$ can not only be directly attended by units from $i-w$ to $i+w$ but has indirectly connected with more units because its immediate neighbors are attended by those units. 

Generally, with the sliding window attention, tokens lose information of a wide range of units in a single step but regain it through depth by stacking multiple layers. With augmenting the depth of layers, a single unit gets increasingly more information.
We can finally get everything in the chunk attended to everything by sliding the small window and stacking the multiple layers. The window size is an engineering trade-off between efficiency and performance. Smaller window sizes are less computationally expensive due to fewer nonzero values (better efficiency), while larger window sizes have richer representation power and often result in performance improvements (higher performance).
The complexity is dominated by $O(w^2 *n)$, where $w$ is window size and $n$ is the length of the sequence chunk. We can simply ignore the constant $w^2$, so the overall complexity is $O(n)$.
As such, Transformer models with the local attention mechanism can handle and process longer input sequences as compared to full-attention Transformers.

\subsubsection{Token Pruning}
\label{sec:pruning}

In this work, with the SparseCoder framework, we will leverage the token pruning approach to prune away tokens layer-wise in Transformer with sparse attention to reduce the computing footprint in the inference stage. 
Learned Token Pruning~\cite{kim2021learned} is adopted in our sparse-attention Transformer, which consists of two token pruning strategies, soft pruning and hard pruning. 
Given a Transformer-based model \textit{M} fine-tuned on security defect detection datasets, the adaptively learned token pruning algorithm comprises three steps as follows:
\begin{itemize}
    \item Step 1: In soft pruning, the model parameters and pruning threshold in each layer are trained by applying the soft mask with decimal masks $\theta_l$, where $\theta_l$ is the soft pruning threshold of layer $l$. 
    \item Step 2: Binarize the decimal mask generated in soft pruning and fix the thresholds, where $S_l(x_i)$ is the importance score of a token $i$ in layer $l$, and $Mask_l(x_i)$ is a binarized mask for token $i$ in layer $l$.
    
    \[
    Mask_l(x_i)= 
    \begin{cases}
        1,& \text{if } S_l(x_i) > \theta_l\\
        0,              & \text{otherwise}
    \end{cases}
    \]
    
    \item Step 3: In hard pruning, remove tokens with binarized masks as 0 and keep the ones with masks as 1. Fine-tune the model parameters after hard pruning.
\end{itemize}

The whole process for token pruning is layer-wise and adaptive. Finally, the token pruning results are visualized in the experimental analysis section to make this black-box methodology more transparent and interpretable to software engineering researchers and engineers.

\begin{figure}[!b]
\centerline{\includegraphics[width=0.7\textwidth,trim = {4cm 1cm 5cm 4cm}, clip]{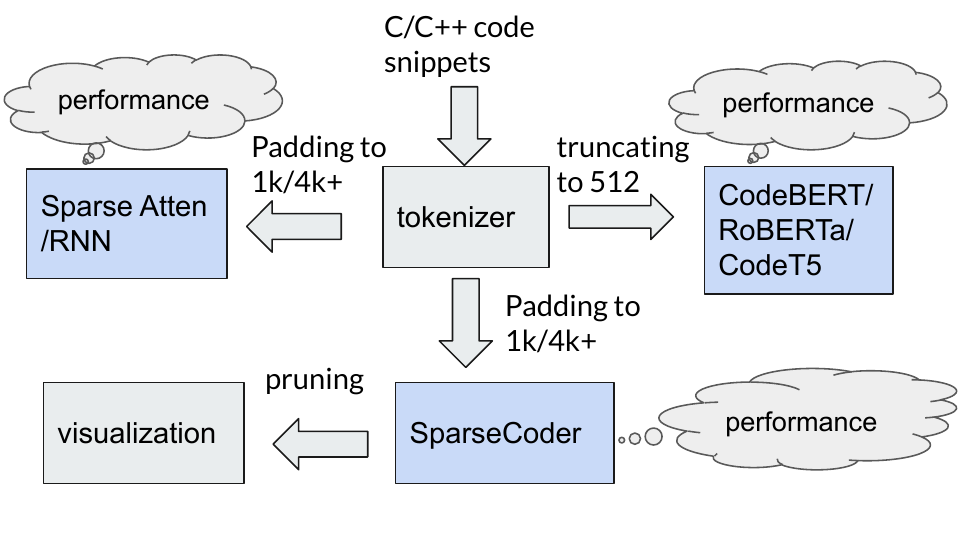}}
\caption{Overview of experimental design of this work. The inputs are C/C++ code snippets in function-level. Shorter sequences than \textit{max\_length} are padded to \textit{max\_length} while longer ones are truncated to the \textit{max\_length}. The \textit{max\_length} for CodeBERT/RoBERTa/CodeT5 is 512, while over 1 or 4 thousands for RNN/Transformer with sparse attention and SparseCoder. (In our experiment, we report the results for \textit{max\_length} as 1024 for RNN/Sparse Atten and SparseCoder since 95\% of our input sequence length is within 1024.).}
\label{fig:experiment}
\end{figure}

\section{Empirical Study of Vulnerability Detection}
\label{sec:experiment}

\subsection{Experimental Design}

The overall design of our systems is shown in Figure~\ref{fig:experiment}.
We apply the logical \textit{OR} operation by combining the positive labels from the multi-task classification with the five Common Weakness Enumerations (CWEs) labels to generate only one label, where the code snippet in function level will be labeled as positive or buggy if at least one of the five types of CWE issues is identified in the specific function.


After that, random downsampling is leveraged to balance the ratio of majority samples which are labeled as non-anomalous in the training set. The testing set is kept unchanged to make a fair comparison in the whole experiment pipeline. As a simple sampling strategy, downsampling is widely used to tackle the imbalance problem, which can also help to reduce the overall training overhead.

We also compare the model efficiency for different models with modified configurations by measuring FLOPs. 
To compare the model efficiency between full attention and sparse attention mechanism, we calculate the FLOPs of different Transformer models by breaking down the Transformer model into FFN (feed-forward layers), Projection Layer (for queries, keys, values and attention outputs), attention layers (full attention or sparse attention) and other operations (e,g., embedding, normalization and multi-heads)~\cite{clark2020electra}. More specifically, we report the FLOPs and performance changes with varying window sizes and sequence lengths.


\subsection{Dataset}

As outlined above in Section~\ref{sec:background}, most of the current works focus on short sequence analysis of programming languages~\cite{wu2020code,wan2018improving,hu2018deep}, likely limited by the capabilities of their algorithms.

Upon conducting a comprehensive literature review on source code analysis and vulnerability detection, we identified a singular dataset containing long sequences that examined:
 1) classification tasks related to source code analysis and vulnerability detection, and 
 2) the reliability of open-source code.
This dataset, introduced by Russell et al.~\cite{russell2018automated} in 2018, delves into vulnerability detection within source code. Notably, the dataset is curated at the function level—the most granular level that provides a comprehensive view of the subroutine flows within the code.

\subsubsection{Data Curation}

Russell et al.~\cite{russell2018automated} compile millions of function-level examples of C/C++ source code snippets from the SATE IV Juliet
Test Suite~\footnote{\url{https://samate.nist.gov/SARD/test-suites/112}}, Debian Linux distribution~\footnote{\url{https://www.debian.org/}}, and public Git repositories on GitHub~\footnote{\url{https://github.com/}}. 
Although the SATE IV dataset has labeled samples with anomalies from 118 different Common Weakness Enumeration (CWE), it consists of synthetic code snippets instead of the original source code, which may not be sufficient as the training set. Debian, known as Debian GNU/Linux, is a free and open-source operating system (OS) with a Linux code basis established by Ian Murdock's team in 1993 and widely applied in many systems. There exists a very well-managed and curated source code of Debian package releases.  
GitHub provides the distributed version control of Git, access control, bug tracking, task management, continuous integration, etc. As of June 2022, a statistical analysis shows that there are over 83 million developers and more than 200 million repositories in GitHub. Compared to Debian packages, Github has a wider range of codebases but is often of lower quality.
Both the samples collected from Debian and Github required extra labeling efforts.

Another essential step of data curation is data cleaning. In open-source projects, code cloning at the function level is commonly observed within or across projects. Similar functions can exist both in training and test sets, although these functions may seem quite diverse at the raw source code level. Removing the potential function duplication can efficiently avoid performance inflation and conceal overfitting caused by such a data leakage issue. Russell et al.~\cite{russell2018automated} conduct a rigorous duplicate removal process via removing functions similar in their lexed representations and feature vectors at the compile level. After the removal process, only 10.8\% of samples pass as distinct functions without duplication and will be utilized in further study.

\subsubsection{Ground Truth}

\begin{figure}[!b]
\centerline{\includegraphics[width=0.6\textwidth,trim = {0cm 0.1cm 1.6cm 1.4cm}, clip]{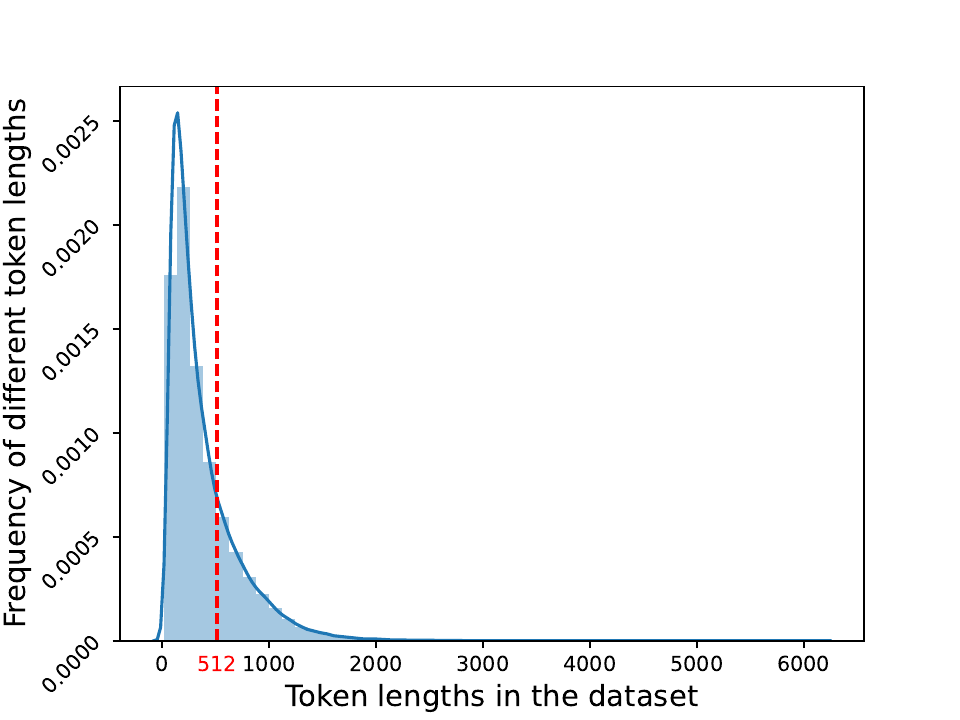}}
\caption{Token length distribution statistic on the test dataset (following the long-tail distribution). As shown in the distribution statistic, over 24\% of our input sequence length is greater than 512 and about 95\% of our input sequence length is within 1024.}
\label{fig:token_len}
\end{figure}


Russell et al.~\cite{russell2018automated} explored the feasibility of 
three approaches, namely static analysis, dynamic analysis, and commit-message/bug-report tagging, to label the collected dataset. However, dynamic analysis is highly computationally expensive, requiring nearly a day of effort to conduct dynamic analysis on 400 functions in ManyBugs Dataset via a single module of the LibTIFF package. A commit-message based approach turns out to be challenging, which cannot guarantee the quality of the label generated and requires extra human efforts to inspect. Finally, only three static analysis tools (namely, Clang~\footnote{\url{https://clang.llvm.org/}}, Cppcheck~\footnote{\url{https://cppcheck.sourceforge.io/}}, and Flawfinder~\footnote{\url{https://dwheeler.com/flawfinder/}}) are used to obtain the ground truth for this dataset. Those different static analysis tools address and detect different aspects of anomalies for C/C++ source code. Clang includes a wide scope of vulnerability detection and additionally checks the programming style, syntax, and other aspects, which are less presumable to be anomalous. Cppchecker provides filename, line, severity, alert identifier (with a message), and CWE for each alert instead of style. Flawfinder utilizes simple text pattern matching and ignores comments and strings, which is geared toward the CWEs instead of style. The multiple analysis results are incorporated and irrelevant vulnerabilities which are not associated with security anomalies are pruned away. A professional security research team generates the binary labels and categorizes the anomalies into five multiple labels, which are summarized in Table~\ref{table:dataset}. In addition, the overall dataset is highly imbalanced. Namely, function-level code snippets identified as at least one of the five CWE issues are less than 40\% as illustrated in Table~\ref{table:dataset}.

We also analyze the tokenization results to see the distribution of token length and the necessity to use SparseCoder. There are over 24 percent samples over the token limitation (512) in the training set. As illustrated in Figure~\ref{fig:token_len}, the distribution of the token length follows the long tail distribution. As shown in the distribution statistic, over 24\% of our input sequence length is greater than 512 and about 95\% of our input sequence length is within 1024. For samples with a token length less or greater than the configuration of maximum token length, we have two schemes padding or truncation to make the input sequence with the same token length as required in Transformer-based models. After repeating the tokenization and visualization process, we find that the test set has the same token length distribution as the training set.

\begin{table}[]
\small
\caption{Ground truth summarization of the security dataset. In our original dataset, each function-level code snippet is label with multiple classes (five CWE types). And it's observed that the target classes are imbalanced, with each class less than 40\% and even no more than 10\% for some classes.}
\begin{tabular}{@{}cll@{}}
\toprule
\textbf{CWE Types}                                                   & \textbf{Anomaly Description}                                                                                                                                    & \textbf{Frequency/\%} \\ \midrule
120/121/122                                                 & Buffer Overflow                                                                                                                                        & 38.2\%       \\
119                                                         & \begin{tabular}[c]{@{}l@{}}Improper Restriction of Operations \\ within the Bounds of a Memory Buffer\end{tabular}                                     & 18.9\%       \\
476                                                         & Null Pointer Dereference                                                                                                                               & 9.5\%        \\
469                                                         & \begin{tabular}[c]{@{}l@{}}Use of Pointer Subtraction to\\ Determine Size\end{tabular}                                                                 & 2.0\%        \\
\begin{tabular}[c]{@{}c@{}}20, 457, 805\\ etc.\end{tabular} & \begin{tabular}[c]{@{}l@{}}Improper Input Validation, Use of \\ Uninitialized Variable, Buffer Access\\ with Incorrect Length Value, etc.\end{tabular} & 31.4\%       \\ \bottomrule
\end{tabular}
\label{table:dataset}
\end{table}

\subsection{Evaluation Metrics}
\label{sec:metrics}

To effectively address the open issues in the classification task of vulnerability detection, it's essential first to establish a clear framework for evaluation. This involves defining key metrics for assessment. Let's consider the outcomes identified by a classifier as illustrated in Table~\ref{table:confusion_matrix}: True Negatives (TN), False Negatives (FN), False Positives (FP), and True Positives (TP).  
Each of these represents a different type of classification result, and understanding their implications is crucial for evaluating the performance of the classifier.

The experimental results are reported in terms of the following metrics: 
precision, recall, F1, false alarm, AUC and loss (namely binary cross-entropy loss~\cite{goodfellow2016deep} on the test set). Accuracy is easy to understand and interpret and widely reported in SE research papers. However, it fails when the dataset is imbalanced or the costs of prediction errors for distinct classes is different. For instance, in our vulnerability detection task (involving imbalanced data where the minority class is positive or code snippets detected with vulnerabilities), it is of greater importance to evaluate whether the model accurately classifies the positive samples. Therefore, we report results in precision, recall, F1, false alarm, AUC to provide a more nuanced view of model performance for imbalanced data~\cite{yang2021learning}.


\begin{table}[ht]
\center
\renewcommand\arraystretch{1.5}
\setlength\tabcolsep{0pt}
\begin{tabular}{c >{\bfseries}r @{\hspace{0.7em}}c @{\hspace{0.4em}}c @{\hspace{0.7em}}l}
  \multirow{2}{*}{\rotatebox{90}{\parbox{6cm}{\bfseries\centering True Label}}} & 
    & \multicolumn{2}{c}{\bfseries Predicted Label} & \\
  &  & \MyBox{True}{Positive} & \MyBox{False}{Negative} & Positive \\[2.4em]
  &  & \MyBox{False}{Positive} & \MyBox{True}{Negative} & Negative \\
  &  & Positive$'$ & Negative$'$ & Total
\end{tabular}
\caption{Confusion matrix for a binary classification task, where true label (positive and negative) equals predicted label (positive$'$ and negative$'$). In our specific context, positives refer to function-level code snippets that contain at least one of the five specified CWE issues, while negatives represent code snippets devoid of any of these five CWE issues.}
\label{table:confusion_matrix}
\end{table}



\begin{itemize}
    \item Precision = TP/(TP+FP), indicating how many of the functions predicted as positive or buggy (containing at least one CWE issue) are actually positive or buggy.
    \item Recall = TP/(TP+FN), the proportion of actual positive cases (functions with at least one CWE issue) that the model correctly identifies as positive.
    \item F1 = 2*TP/(2*TP+FP+FN), the harmonic mean of precision and recall of the classification model.
    \item False alarm = FP/(FP+TN), the percentage of the non-buggy functions out of ones that are predicted as buggy.
    \item AUC (Area Under the Curve): a performance metric for binary classification models that measures the ability of the model to distinguish between the positive and negative classes across all possible thresholds.
    \item Loss (binary cross-entropy loss~\cite{goodfellow2016deep}) =  $-\frac{1}{N} \sum_{i=1}^{N} [ y_i \cdot \log(p_i) + (1 - y_i) \cdot \log(1 - p_i) ]$, where $N$ is the number of observations, $y_i$ is the actual label, and $p_i$ is the predicted probability. It evaluates the performance of a classification model whose output is a probability value indicating how likely it is that a given input sample belongs to a positive class. 
\end{itemize}

We further report the model efficiency measured in FLOPs (floating point operations).  
Previous research works~\cite{chen2023longlora,kim2021learned} have demonstrated that FLOPs as a standard measure agnostic to the hardware performance, indicating how many floating point calculations it performs in a single second. Training and inference in neural networks, especially deep learning models, involve a vast number of matrix multiplications and other operations.  FLOPs provide a standardized  way to estimate and compare the computational effort required by different models or frameworks.
As LLMs grow in size, efficiency becomes paramount. These models, with their billions or even trillions of parameters, demand significant time and energy resources during both pre-training and deployment. Therefore, measuring and optimizing based on FLOPs is essential for model efficiency and model scalability in practical software engineering deployments.



\subsection{Statistical Tests}\label{tion:stats}

In this study, we report the median results of repeated ten runs for each group of experiments.
To select ``best'' learning methods, we follow the advice of Rosenthal et al.~\cite{rosenthal1994parametric} by conducting statistical tests. 
Rosenthal et al. discuss different parametric methods for asserting that one result is with some small effect of another (i.e. it is ``close to'').
They list dozens of effect size tests that are divided into two groups: the $r$ group that is based on the Pearson correlation coefficient; or the $d$ family that is based on absolute differences normalized by the size of the standard deviation. By utilizing the most direct $d$ family method, it can be concluded that one distribution is the same as another if their mean value differs by less than Cohen's delta ($d$*standard deviation), where $d$ is computed separately for each different evaluation measure~(
precision, recall, F1, false alarm, AUC and loss).
 
To visualize that ``close to'' analysis, in all our results:
\begin{itemize}
    \item We calculate the standard deviation of each row in Table~\ref{table:all_results}, Table~\ref{table:sequence_len} and Table~\ref{table:window_size} which formulated as $STDEV$
    \item Any cell that is within $d*STDEV$ of the best value will be highlighted in \textcolor{pink}{red} or \textcolor{gray}{gray}. All \textcolor{pink}{red} cells are observed as ``winners'' to maximize and all the \textcolor{gray}{gray} cells are ``winners'' for the rows to minimize. The other cells without highlighting are ``losers''.
    \item For 
    precision, recall, F1 and AUC, the ``best'' cells have ``highest value'' as \textcolor{pink}{red} since the optimization goal is to maximize these values. For false alarm and loss, the ``best'' cells have the ``lowest value'' marked as \textcolor{gray}{gray} since those metrics are to be minimized. 
\end{itemize}

We follow the advice of
a widely cited paper by
Sawilowsky~\cite{sawilowsky2009new} as a standard when deciding the value of $d$ in our statistical analysis, which
asserts that ``small'' and ``medium'' effects can be measured using $d=0.2$ and $d=0.5$ (respectively).
Splitting the difference, we will analyze this data by looking for differences larger than $d=(0.5+0.2)/2=0.35$.

\subsection{Extractor of Attention Scores in Sparse Attention Matrix}

Inspired by LTP, which was originally proposed based on I-BERT~\cite{kim2021bert} (a variant of RoBERTa with full attention mechanism). To develop SparseCoder, we further modified the existing implementation to adapt the token pruning algorithm for Transformer with sparse attention. This adaption was necessary not only because the implementation of LTP's implementation is closely tied to the I-BERT framework, which isn't a modular component, but also due to the need to accommodate sparse attention within LTP. 
As depicted in Figure~\ref{fig:attentions}, consider a short sequence of ten tokens, with two special tokens (the 1st and 8th). Given a window size of 3, the sparse attention matrix is represented in Sub-figure (b), with global attention highlighted in red and local attention in green. The attention matrix from Sub-figure (b) is decomposed into global attention in Sub-figure (c) and local attention in Sub-figure (d) for clearer representation.
For the special tokens in global attention, they attend to every other token in the sequence.

In our implementation, given the local attention matrix is sparse, we avoid storing the attention scores in a $n^2$ matrix format. Instead, the sparse attention matrix is reshaped into a $n*(2w-1)$ matrix to economize on storage space. In Sub-figure (d), the gray cells indicate no-attention areas, while the dark green cells represent attention scores to themselves. 
We accumulate the sub-diagonals of the reshaped local attention score matrix to derive the importance score of each token within local attention. For instance, the local attention score for the 3-rd token is revealed by the sub-diagonal in Sub-figure (d).
Subsequently, the global attention matrix for special tokens is aggregated with local attention scores of the corresponding special tokens. This provides a comprehensive importance score for each token in each layer.

\section{Experimental Results}
\label{sec:result}

In this section, we will answer the research questions raised above. We also compare different model efficiency with modified configurations by reporting FLOPs during model inference, which is widely utilized as a standard measure of model efficiency to get real-time prediction and bring the  processors of large models to the edge~\cite{gholami2021survey,clark2020electra}. All of the models are fine-tuned based on released checkpoints from Hugging Face~\footnote{\url{https://huggingface.co/}} (a community and data science platform that provides standard tools to enable users to build, train and deploy ML models based on open-source code and technologies). Our reported experiment results are generated with a fixed training set and tested on the same test set generated from the data curation discussed in Section~\ref{sec:experiment}.

\subsection{RQ1}
\begin{RQ}
{\bf RQ1.} \textbf{\textit{Scalability:}} {\em Does sparse attention mechanism advance the model scalability compared with the benchmark models (RNN-based method and Transformer with full attention, namely RoBERTa, CodeBERT and CodeT5)?}
\end{RQ}

To answer this research question, we conduct experiment with SparseCoder, Sparse Atten, RoBERTa, CodeBERT, CodeT5 and RNN model on our security vulnerabilty dataset. As stated in the Section~\ref{sec:introduction}, most of prior works focus on short sequence analysis fail in analyzing longer input sequences as the source code snippets under test get increasingly sophisticated. Therefore, it's importance to study which framework can scale and address the long seqence analysis.

For the configuration of the recurrent neural network, the maximum sequence length is set as 1024, which is the same setting as the \textit{max\_length} in Transformer with sparse attention in our experiments. For instance, shorter sequences than 1024 are padded while longer ones are truncated to the \textit{max\_length}. After the function-level code snippets are converted to a numeric look-up table by an embedding layer with the integer-encoded vocabulary, a convolutional layer is utilized to extract the underlying features of the embedding matrix, where the filter size is set as 512 (same as the maximum window size of Transformer with sparse attention and configs in RoBERTa and CodeBERT considering the full attention mechanism). Subsequently, the number of middle layers with gated recurrent units is configured as 12, since the default number of layers in the Transformer-based model is 12, to make the comparison fair. 
After 12 layers of GRU, the max pooling layer is leveraged to down-sample the input representation, followed by three dense layers to fully connect the network and to make a binary prediction on whether the given sequence of code snippets is vulnerable or not.

Our experiments show that SparseCoder, Sparse Atten and RNN model can all solve longer sequences with token length longer than 512, where in our experiment we set the upper bound as 1024 since 95\% of our test samples have sequence length within 1024 as indicated in Figure~\ref{fig:token_len}. While for RoBERTa, CodeBERT, CodeT5, these models fail with max length of input sequences greater than 512. As indicated in Table~\ref{table:all_results}, the overall performance of RNN model is worse compared with other Transformer-based models. Compared with RoBERTa, CodeBERT and CodeT5, it can be concluded that sparse attention mechanism can advance the model scalability of Transformer-based models by extending the in context length that models can analysis. In the following research questions, we would further analysis from the aspects of efficiency, performance and etc.




\begin{figure} 
\centerline{\includegraphics[width=0.6\textwidth,trim = {0cm 0cm 1cm 0.5cm}, clip]{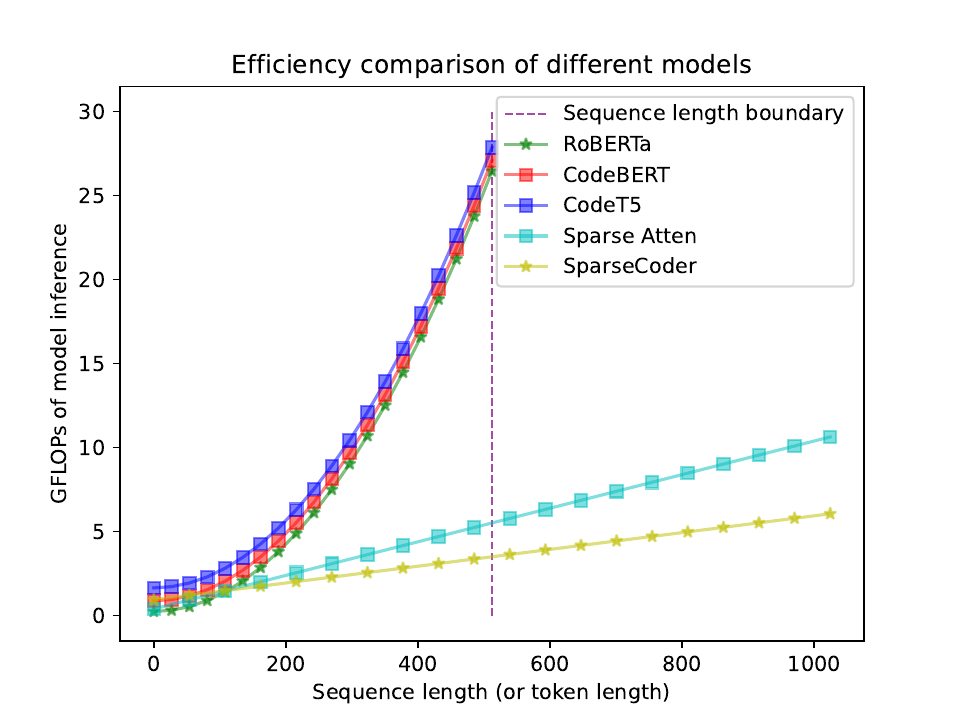}}
\caption{Comparing GFLOPs of different models in the inference stage, where \(1 \text{ GFLOPs } = 10^9 \text{ FLOPs}\). The FLOPs simulation is based on Electra~\cite{clark2020electra}. The dashed line denote the sequence length boundary as 512 which reported by previous research works~\cite{vaswani2017attention,beltagy2020longformer}. With the increasing sequence length, the GFLOPs grows quadratically for RoBERTa, CodeBERT and CodeT5, while linearly for Sparse Atten and SparseCoder. Overall, SparseCoder requires a lower GFLOPs compare to Sparse Atten.}
\label{fig:efficiency}
\end{figure} 

\subsection{RQ2}
\begin{RQ}
{\bf RQ2.} \textbf{\textit{Efficiency:}} {\em How does sparse attention mechanism advance the model efficiency compared with the benchmark models (Transformer with full attention, namely RoBERTa, CodeBERT and CodeT5)? }
\end{RQ}

To answer this research question, we simulate GFLOPs of each model in the inference stage based on a prior study, Electra~\cite{clark2020electra}. Moreover, multiple studies~\cite{clark2020electra,chen2023longlora,kim2021learned} report floating point operations (FLOPs) to estimate and compare the computational effort required by different models or frameworks since it's considered as a
measure agnostic to the particular hardware for low-level optimizations.
As a mathematical computation, for element-wise multiplication of two matrices of size $m*n$, we count matrix multiplications as $2*m*n$ FLOPs instead of $m*n$, which one might consider if using fused multiply-add operations. This process involves breaking down the mathematical operations (including multiplications, additions and other operations) performed at each layer and summing them up.

Also, we report the model inference time for different frameworks (measured on a fixed subset of one thousand testing samples) and we rank the model performance in order of incrementing model inference
times from left to right. Table~\ref{table:all_results} shows that SparseCoder is satisfactory with respect to predictive performance while least inference time and achieves similar performance as illustrated by statistic test (only less than 1\% performance drop in terms of precision, F1, false alarm, AUC and loss.)
In source code analysis, decreasing the cost of inspecting falsely reported warnings generated by static code analysis tools is crucial for software engineers (especially in the early stage of a software project's life cycle) and provides a meaningful guideline to improve the performance of current SA tools~\cite{yang2021understanding}. Moreover, other Transformer-based models (RoBERTa, CodeBERT and CodeT5) also outperform the RNN model.
Although our RNN model can conduct inference on longer sequences than traditional Transformer models (RoBERTa and CodeBERT), we can conclude that Transformer-based models are more prominent than RNN-based models in long-term dependency code analysis.


Note that the Transformer with sparse attention is pre-trained based on the checkpoint of RoBERTa and both these two models are pre-trained on a large volume of the document corpus.
Compared with CodeBERT and CodeT5 (which pre-trained on programming languages), Transformer with sparse attention only performs better than CodeBERT in terms of 
false alarm slightly, and worse than CodeBERT in terms of precision, recall, F1, AUC and loss. 
In terms of model efficiency measured in FLOPs, it's observed that a Transformer with sparse attention significantly reduces the computational cost in the inference stage compared with RoBERTa, CodeBERT and CodeT5. Also, as demonstrated above, Transformer models with full attention have restricted capability to analyze long sequences. Considering that our sparse attention Transformer checkpoint was pre-trained on NLP datasets, we are optimistic about its potential to yield improved outcomes once pre-trained on SE datasets for industrial use cases.\\
\begin{table}[]
\small
\caption{Experiment results (median of ten runs) of RNN-based model and Transformer-based models, i.e., RoBERTa, CodeBERT, CodeT5, Transformer with sparse attention and SparseCoder. The different frameworks are ranked} by the incremental model inference times measured on 1k samples from left to right. ( ``Sparse Atten'' refers to Transformer model incorporating with sparse attention mechanism (without the learned token pruning), while in ``SparseCoder'' we further extend the framework with a learned token pruning algorithm to adaptively prune away trial tokens layer-wisedly during the model fine-tuning and inference stage.)
Note that SparseCoder performs arguable as good as anything else while at the same time, consumes far less GPU resources (see Figure~\ref{fig:efficiency}).
In this figure, pink denotes results
that are significantly larger (and better) than other models while gray denotes results that are significantly smaller (and better).
\tabcolsep=0.11cm
\begin{adjustbox}{max width=1.0\textwidth}
~~~~~~~\begin{tabular}{l|llllll}
\toprule
\multicolumn{1}{c}{} & \multicolumn{1}{c}{Sparse-} & \multicolumn{1}{c}{} & \multicolumn{1}{c}{} & \multicolumn{1}{c}{} & \multicolumn{1}{c}{} & \multicolumn{1}{c}{}\\ 
\multicolumn{1}{c}{\multirow{-2}{*}{\textbf{Metrics}}} &Coder & \multicolumn{1}{c}{\multirow{-2}{*}{RNN}} &\multicolumn{1}{c}{\multirow{-2}{*}{Sparse Atten}} & \multicolumn{1}{c}{\multirow{-2}{*}{RoBERTa}} & \multicolumn{1}{c}{\multirow{-2}{*}{CodeBERT}} & \multicolumn{1}{c}{\multirow{-2}{*}{CodeT5}} \\ 
\midrule
Precision & \red 29.56 & 15.36  & \red 30.12 &  \red 29.17 & \red 30.12 & \red 30.31   \\
Recall & 78.64 & 79.49  & 78.82 & 77.91 & \red 80.13 & \red 81.95  \\
F1 & \red 43.14 & 25.75 & \red 43.59 & \red 42.45 & \red 43.79 & \red 44.26   \\
\textbf{False alarm} & \gray 12.98 & 30.33 & \gray 12.66 & \gray 13.10 & \gray 12.87 & \gray 13.05   \\
AUC & \red 89.12 & 82.46 & \red 89.34 & 88.99 & \red 90.08 & \red 89.87 \\
Loss & \gray 0.412 & 0.678  & \gray 0.390 & \gray 0.391 & \gray 0.373 & 0.502  \\
Time & \gray 4.54 & \gray 6.87 & 14.72  & 17.81 & 17.90  & 18.54  \\
\bottomrule
\end{tabular}
\label{table:all_results}
\end{adjustbox}
\end{table}

\subsection{RQ3}

\begin{table}[]
\small
\caption{Experiment results (median of ten runs) of ablation study via modifying max sequence length in Transformer with sparse attention with fixed window size = min(512, max\_length), where padding and truncation are utilized.}
\tabcolsep=0.11cm
\begin{adjustbox}{max width=1\textwidth}
~~~~~~~~~~\begin{tabular}{lllllll}
\toprule
\multicolumn{1}{c}{} & \multicolumn{6}{c}{Maximum sequence length}  \\ \cmidrule(l){2-7} 
\multicolumn{1}{c}{\multirow{-2}{*}{\textbf{Metrics}}} & n = 32 & n = 64 & n = 128 & n = 256 & n = 512 & n = 1024 \\ \cmidrule(r){1-1}
Precision & 14.03 & 16.02 & 17.38 & 23.44 & \red 28.49 & \red 30.12\\
Recall & 70.91 & 74.35 & 74.53 & 77.40 & \red 78.42 & \red 78.82\\
F1 & 23.43 & 26.36 & 28.18 & 35.99 & \red 41.79 & \red 43.59 \\
\textbf{False alarm} & 30.08 & 26.99 & 24.54 & 17.51 & \gray 13.64 & \gray 12.66 \\
AUC & 77.87 & 81.19 & 82.49 & 87.25 & \red 88.89 & \red 89.34 \\
Loss & 0.559 & 0.528 & 0.523 & 0.432 & \gray 0.410 & \gray 0.386 \\
\bottomrule
\end{tabular} 
\label{table:sequence_len}
\end{adjustbox}
\end{table}

\begin{table}[]
\small
\caption{Experiment results (median of ten runs) of ablation study via modifying window size in Transformer with sparse attention with fixed max length as 1024 with padding and truncation. Run time reported in this table refers to the inference time on 1k samples on the same hardware (GPU). It's observed that the inference time reduces with the decreasing window sizes.}
\tabcolsep=0.11cm
\begin{adjustbox}{max width=1\textwidth}
~~~~~~~~~~~\begin{tabular}{lllllll}
\toprule
\multicolumn{1}{c}{} & \multicolumn{6}{c}{Window size}  \\ \cmidrule(l){2-7} 
\multicolumn{1}{c}{\multirow{-2}{*}{\textbf{Metrics}}} & w = 16 & w = 32 & w = 64 & w = 128 & w = 256 & w = 512 \\ \cmidrule(r){1-1}
Precision & 28.79 & 28.41 & 28.83 & 28.35 & 29.70 & \red 30.12\\
Recall & 77.98 & 78.25 & 80.01 & 79.10 & 77.95 & \red 78.82\\
F1 & 42.05 & 41.68 & 42.38 & 41.74 & 43.25 & \red 43.59 \\
False alarm & 13.36 & 13.66 & 13.68 & 13.85 & 13.03 & \gray 12.66 \\
AUC & 88.45 & 88.64 & \red 89.52 & 88.68 & 89.23 & 89.34 \\
Loss & 0.421 & 0.414 & 0.397 & 0.409 & \gray 0.390 & \gray 0.386 \\
Time & \gray 9.98 & 10.7 & 10.68 & 11.03 & 12.03 & 14.48 \\
\bottomrule
\end{tabular}
\label{table:window_size}
\end{adjustbox}
\end{table}



\begin{RQ}
{\bf RQ3.}
{\em How do the modified window size  and the token length impact the performance of the Transformer with a sparse attention mechanism?}
\end{RQ}

To answer this research question, we conduct the ablation study to explore the impact of modified window size on the performance of the Transformer with sparse attention. For each group of the experiment, the window size is set as $\{16, 32, 64, 128, 256, 512\}$ and the maximum sequence length is set as 1024 as a control variable. As demonstrated in Table~\ref{table:window_size}, given the maximum sequence length as a constant, the overall performance gets slightly better as the window size grows. Moreover, FLOPs in the model inference stage are significantly reduced by configuring a smaller window size. This illustrates that local attention is efficient in reducing the memory requirement without damaging the model performance significantly. Based on this observation, we suggest utilizing the sparse attention mechanism when applying Transformer-based models to software engineering tasks. 

We also conduct the ablation study with a modified maximum sequence length on the Transformer with sparse attention to explore the impact of the different sequence lengths on the model performance. For each group of the experiment reported in Table~\ref{table:sequence_len}, the maximum sequence lengths are set as $\{32, 64, 128, 256, 512, 1024\}$ respectively. Since the window size cannot be greater than the sequence length and it also has the 512 constraint from the inherent nature of full attention, the window size would be $min \{ 512, max\_length\}$, which will be set as $\{32, 64, 128, 256, 512, 512\}$ correspondingly. As illustrated in Table~\ref{table:sequence_len}, the overall performance of Transformer with the sparse attention gets improved as the maximum sequence length is augmented, which indicates longer code sequences with less truncation are a benefit for the model performance. 

\subsection{RQ4}

\begin{RQ}
{\bf RQ4.}
\textbf{\textit{Interpretability:}} {\em Can SparseCoder advance interpretability of the Transformer-based models via a token pruning algorithm?}
\end{RQ}




To answer this research question, we further extend Sparse Atten (Transformer with only sparse attention mechanism) to SparseCoder via incroprating an adaptively learned token pruning algorithm.
Generally, when conducting token pruning, there exists a trade-off between the pruning threshold and model performance. A higher pruning threshold, which removes more tokens, will reduce computing overhead but with a dropped model performance. Therefore, we conduct an ablation study to decide the optimal configuration of a final layer as 0.01. For each of the remaining layers, the threshold will be linearly scaled, for example, it's set as 0.0025, 0.005, and 0.0075 respectively for the 3rd, 6th, and 9th layers when the threshold of the final layer (12-th) is configured as 0.01.

\begin{figure} 
\centerline{\includegraphics[width=0.45\textwidth,trim = {3cm 17.5cm 11cm 3.7cm}, clip]{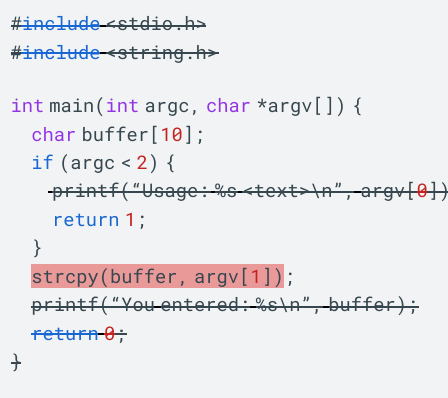}}
\caption{A demo of visualization of SparseCoder on C code snippet for the security defect detection. As illustrated, the line of code highlighted in red introduces a buffer overflow issue.}\label{fig:visualize_code}
\end{figure} 

Figure~\ref{fig:visualize_code} presents a visualization demo of security defect detection based on function-level C code. Strikethroughs in the code snippet highlight the trivial information that the algorithm pruned; this information was subsequently ignored by SparseCoder during the fine-tuning and testing phases. After pruning, the underlying buffer overflow issue, emphasized in red, becomes more readily discernible. Through such token pruning, our proposed SparseCoder not only enhances interpretability through visualization but also improves efficiency, as measured in GFLOPs, advanced by about \textbf{two times} comparing with sparse Transformer without token pruning (Sparse Atten), as depicted in Figure~\ref{fig:efficiency}. 
Also, as a trade-off of the advanced efficiency, only less than 1\% performance drop on the 
precision, recall, F1, false alarm (lower the better) and AUC
is observed in Table~\ref{table:all_results}. This advancement notably facilitates real-time analysis and model interpretability.



\section{Discussion}
\label{sec:discussion}

\subsection{Threats to Validity}
\label{sec:threats}


Threats to validity that may threaten the generality of our conclusions drawn in this paper with datasets found in the future are listed as follows:

\textbf{Sampling bias.}
Regarding sampling bias, our first comment is that the conclusion drawn in this paper is based on the dataset explored in this specific empirical study, security defect detection on open-source C/C++ projects. In future work, it's necessary to repeat these experiment rigs on a new dataset composed of other programming languages or other downstream software engineering tasks to verify the generality of this framework.

Besides, in this experiment, random sampling is utilized to advance the computational efficiency and to balance the binary labels in the training set. The negative samples in the training set were randomly down-sampled with a fixed random seed and the dataset after down-sampling was dumped into a CSV file to make the experiment repeatable and comparable. However, the down-sampling process might still incur some biased issues such as losing information from some important instances.


\textbf{Parameter bias.}
In this work, there exist multiple parameters to configure in each classification model leveraged in this work, namely, the number of layers in the RNN model, learning rate and batch size in RoBERTa and CodeBERT and window size in Transformer with sparse attention and SparseCoder. Model performance might be improved by tuning these configurations. However, this study emphasizes more on model parameter optimization. And the batch size of each model is set to as much as the computing resource can process. And the learning rate of each model is configured as the same value to make a fair comparison. For window size in Transformer with sparse attention, we conduct a series of ablation studies with window size configured as $2^n$ where $n = 5,6,7,8,9,10$. In the Transformer with sparse attention, the author suggests a scheme of small window size in lower layers to capture local information and large window size in higher layers to represent the high-level or wholesome information of the sequence. These configurations can also be an influencing factor in the experiment results.

\subsection{Future Work}
\label{sec:future_work}
 

In this subsection, we outlook the potential directions of future work based on the framework explored in this study. 

The first potential direction involves pre-training SparseCoder with programming language datasets to enhance the model's ability to understand the programming language.  
A notable instance is CodeBert, pre-trained on the combination of natural language datasets and six programming datasets based on the checkpoint of RoBERTa and achieving superior performance in various downstream tasks compared to state-of-the-art models. 
However, given the constraints on computational resources at our academic institution, we will not focus extensively on this path. This area of research is more suited to major tech corporations equipped with extensive high-performance computing facilities.

The second prospective research tendency is incorporating the inherent structure of programming language in the training and evaluation of specific models. In contrast to natural language datasets consisting of sequential data flow, programming language datasets exhibit more structural complexity and grammatical stringency. 
Many of current applications of NLP models on programming language tasks simply consider the source code as a sequence of a kind of language, instead of taking account of the structure of the source code. Although there exist research works~\cite{kim2021code,chirkova2021empirical} about combining the code structure (such as AST or graph-based) into the embedding of such Transformer-based models, a systematic study is requisite to draw a solid conclusion about the comparison between sequence or code structure on source code analysis. 

The third future direction is incorporating the modules (sparse attention mechanism and learned token pruning algorithm) proposed in SparseCoder to other large language models.
Theoretically, our proposed combination of sparse attentions and learned token pruning could also be applied to other LLMs since the sparse attention and token pruning can be adopted to LLMs with self-attention mechanism. It's capable to be applied to existing models (either encoder or decoder models, e.g., Lora, T5 and GPTs), and we find more recent work LongLora~\cite{chen2023longlora} and Infini-attention~\cite{munkhdalai2024leave} from Google. Since these two technologies are not encapsulated independently from the LLM, applying them requires modifications to both LLMs' standard architecture (as we replaced the attention module from self-attention to sparse attention and updated the token importance scores with an adaptively learned threshold to the standard Longformer codebase/libirary). Similar works and explorations, such as LongLora~\cite{chen2023longlora} and Infini-attention~\cite{munkhdalai2024leave}, indicate that extending the sequence/context length that LLMs can analysis is a promising research topic that researchers (not limited to SE domain) are continuously working torwards.

Finally, assessing the applicability of SparseCoder across various downstream software engineering tasks is crucial.
The framework was also adapted to an additional security dataset (PowerShell dataset derived from Windows Defender system) during an internship in Microsoft Research. However, due to data confidentiality concerns, results from the ablation study and details about the proprietary dataset will remain unpublished and not open-source.
In lieu of this, substantial effort in our work has been channeled into identifying a high-quality public dataset that contains labeled raw source code. Current trends in the Software Engineering community largely focus on analyzing short code sequences, typically constrained to approximately 150-400 tokens, sometimes even less, as seen in studies~\cite{hu2018deep,wan2018improving,wu2020code,zhang2020retrieval}. The process of collecting and labeling data, especially when sifting through extensive repositories of raw source code, is extremely resource-demanding. This creates significant obstacles in procuring suitable datasets or case studies (especially those involving long sequence analysis) from public or academic sources for our empirical study.

\section{Conclusion}
\label{sec:conclusion}
This paper introduces SparseCoder, a novel Transformer-based approach for identifying security vulnerabilities in C/C++ source code.
SparseCoder stands out due to its integration of sparse attention mechanisms and an adaptive token pruning algorithm within Transformer-based models.\\
Our experiments, conducted on two hundred thousand data points sampled from one billion function-level code snippets in open-source C/C++ projects,
highlight three key advantages of SparseCoder:
\begin{itemize}  
\item {\em \textbf{Scalability}}:
As demonstrated in Table~\ref{table:all_results} and Table~\ref{table:sequence_len}, SparseCoder can handle longer input sequences than the prior state-of-the-art models.
\item {\em \textbf{Efficiency}}:
As shown in Table~\ref{table:all_results} and Figure~\ref{fig:efficiency}, SparseCoder operates significantly faster than other SOTA methods while maintaining comparable performance. Our statistical tests reveal a minimal performance drop (less than 1\%), and SparseCoder scales linearly in terms of hardware requirements measured in FLOPs, unlike other methods, including the state-of-the-art models (RoBERTa, CodeBERT, and CodeT5), which scale quadratically as input sequence length increases.
\item {\em \textbf{Interpretability}}:
Figure~\ref{fig:visualize_code} illustrates how SparseCoder enhances interpretability by visualizing non-trivial tokens layer-wise. This feature aids software developers in pinpointing segments that potentially cause bugs, rather than merely providing a binary prediction.
\end{itemize}

\section{Data Availability Statements}

To facilitate further work by other researchers in this area, all of our scripts and datasets are available on-line\footnote{ \url{https://github.com/invisiblehead/Sparse_Attention_on_Transformer-based_model}.}. 

\section{Conflict of Interest}

The authors declared that they have no conflict of interest.

\balance

\bibliographystyle{spmpsci}      

\bibliography{main}

\end{document}